\documentclass[a4paper,fleqn,usenatbib]{mnras}
\usepackage[T1]{fontenc}
\usepackage{ae,aecompl}
\usepackage{graphicx}	
\usepackage{amssymb}
\usepackage{amsmath}	
\usepackage{times}

\def\eq#1{{Eq.~(\ref{#1})}}

\newcommand{\HI}{HI}

\title[CO intensity mapping]{Constraining the CO intensity mapping power spectrum at intermediate redshifts}

\author[Padmanabhan]{Hamsa Padmanabhan\thanks{Email: hamsa.padmanabhan@phys.ethz.ch}
\\
  Institute for Particle Physics and Astrophysics, ETH Zurich, Wolfgang-Pauli-Strasse 27, CH-8093 Z\"{u}rich, Switzerland
  }

\date{Accepted ---. Received ---; in original form ---}

\pubyear{2017}

\begin{document}
\label{firstpage}
\pagerange{\pageref{firstpage}--\pageref{lastpage}}
\maketitle

\begin{abstract}
We compile available constraints on the carbon monoxide (CO) 1-0 luminosity functions and abundances at redshifts 0-3. This is used to develop a data driven halo model for the evolution of the CO galaxy abundances and clustering across intermediate redshifts. It is found that the recent constraints from the CO Power Spectrum Survey \citep[$z \sim 3$;][]{keating2016}, when combined with existing observations of local galaxies \citep[$z \sim 0$;][]{keres2003}, lead to predictions which are consistent with the results of smaller surveys at intermediate redshifts ($z \sim 1-2$). We provide convenient fitting forms for the evolution of the CO luminosity - halo mass relation, and estimates of the mean and uncertainties in the CO power spectrum in the context of future intensity mapping experiments. 
\end{abstract}

\begin{keywords}
diffuse radiation -- large scale structure of universe --  cosmology:theory
\end{keywords}

\section{Introduction}
Understanding the evolution of the carbon monoxide (CO) molecular abundance across redshifts  is important from the point of view of galaxy formation, and the star formation history of the universe \citep[for a recent review, see e.g.,][]{madau2014}. CO is strongly associated with star-forming galaxies \citep{carilli2013}, and is the second most abundant molecular species in the interstellar medium, next to molecular hydrogen (H$_2$). {{The CO luminosity and gas mass of local galaxies are well-correlated with the far-infrared luminosity ($L_{\rm FIR}$) which, in turn,  is an indicator of star formation rate \citep{kennicutt1998}. The CO molecule (unlike H$_2$) has a permanent dipole moment and a `ladder' of states for the rotational transitions, making it an ideal probe of the cold neutral phase of the interstellar medium (ISM). The two brightest CO lines are the (1-0) and (2-1) transitions with frequencies 115 and 230 GHz respectively, which also have overlaps with the frequency bands in CMB observations. CO has been studied both with surveys of local galaxies, e.g. \citet{young1995, helfer2003, leroy2009} as well as in individual systems, e.g. \citet{aravena2012, walter2014}.}}

Intensity mapping, in which it is attempted to image the aggregate emission from several sources over very large volumes, does not require the resolution of individual galaxies. This technique has been successfully used to constrain the abundance and clustering of neutral hydrogen (HI) systems around redshift $z \sim 1$ \citep{chang10, masui13, switzer13}, and is a promising probe of cosmology, large scale structure in the universe and galaxy evolution. {The CO molecule offers interesting prospects for intensity mapping, both at intermediate and high redshifts. Intensity mapping with the CO line provides information about the spatial distribution of star formation.} At high redshifts ($z > 6$), performing a CO intensity mapping survey is expected to lead to valuable insights into the physics of galaxies that reionized the universe. It is also a useful probe of the earliest epochs of star formation activity \citep{carilli2011} and large scale structure \citep{righi2008, mashian2015}. 

At intermediate redshifts ($1 < z < 4$), there are good prospects for detecting CO in intensity mapping surveys, especially around the peak of star formation activity at around $z \sim 2-3$ \citep[e.g.,][]{lilly1996, madau1998}.  Recently, \citet{keating2016} have provided the latest constraints on the CO (1-0) intensity power spectrum from the CO Power Spectrum Survey (COPSS) at $z \sim 3$. Future surveys such as the CO Mapping Array Pathfinder (COMAP)\footnote{http://www.astro.caltech.edu/CRAL/projects.html\#comap} will aim to detect CO in emission over redshifts 2-3. There are also prospects for intensity mapping cross-correlations with the results from other surveys such as with  neutral hydrogen (\HI) and the redshifted \textsc{c ii} (158 $\mu$m) {transition} \citep[e.g.,][]{switzer2017}. With the advent of facilities like the ALMA,\footnote{http://www.almaobservatory.org/} (Atacama Large Millimetre Array) and instruments  on the LMT (Large Millimetre Telescope)\footnote{http://www.lmtgtm.org/}, a large number of CO detections in emission from galaxies will be possible. It will be also possible to observe higher transitions of the CO ladder of states, and cross-correlations of multiple spectral lines originating from the same redshift are expected to be useful in statistically isolating the intensity fluctuations \citep{visbal2010}. 

On the theoretical front, a number of approaches have focused on modelling the intensity mapping signal from various CO transitions at different redshifts, to be detected with current and future facilities. These include simulations and semi-analytical methods (SAMs) of galaxy formation \citep{obreschkow2009, fu2012, li2015} which model the metallicity, atomic and molecular gas evolution in galaxies, as well as more empirical techniques starting from the far infrared (FIR) luminosity function \citep{vallini2016}. {These have been studied both around the reionization epoch \citep[$z \gtrsim 6$;][]{righi2008, visbal2010,lidz2011, gong2011}, and  at the peak of the star formation history \citep[$z \sim 2$;][]{pullen2013, li2015}.} One of the chief astrophysical uncertainties in the modeling of the CO power spectrum comes from the knowledge of the CO luminosity ($L_{\rm CO}$) - host halo mass ($M$) relation. Various functional forms for this relationship have been suggested in the literature, based on the results of forward modelling such as SAMs and hydrodynamical simulations (a recent summary is provided in \citet{li2015}). The theoretical  models are found to lead to predictions spanning over an order of magnitude in the CO power spectrum \citep[e.g.,][]{breysse2014, li2015}.

In this paper, we adopt a complementary, empirical approach, anchored to the observational data, towards understanding the evolution of the $L_{\rm CO} - M$ relationship.  We begin by reviewing (Sec. \ref{sec:formalism})  the standard formalism and ingredients for calculating the CO power spectrum from intensity mapping. We also provide an overview of the theoretical models of CO from  the literature. Using the empirically determined relations between the CO luminosity, star formation rate and host halo mass, we connect the low redshift CO galaxy measurements and the higher-redshift constraints from intensity mapping into an analytical halo model in Sec. \ref{sec:model}.  We  provide convenient fitting forms and estimates on the errors in the derived parameters, that agree well with the observations of the CO luminosity function at intermediate redshifts. The form of the halo model enables ease of comparison to the empirically determined stellar mass-halo mass relation. In Sec. \ref{sec:compare}, we explore the consistency of the approach with the results of previous literature, and estimate the mean and uncertainties in the CO power spectrum to be observed with current and future facilities. We summarize the results and discuss the future outlook in a brief concluding section (Sec. \ref{sec:conc}). Throughout the paper, we use the $\Lambda CDM$ cosmology with the cosmological parameters $h = 0.71, \Omega_m = 0.281, \Omega_b = 0.046, \Omega_{\Lambda} = 0.719, \sigma_8 = 0.8, n_s = 0.963$.

\section{Formalism}
\label{sec:formalism}
In this section, we briefly review the standard equations involved in calculating the CO power spectrum observed in intensity mapping, similar studies are outlined in \citet{breysse2014}, \citet{mashian2015}, \citet{pullen2013}.

The specific intensity of a CO line  observed at a frequency, $\nu_{\rm obs}$ is given by:
\begin{equation}
 I(\nu_{\rm obs}) = \frac{c}{4 \pi} \int_0^{\infty} dz' \frac{\epsilon[\nu_{\rm obs} (1 + z')]}{H(z') (1 + z')^4}
\end{equation} 
in which $H(z)$ is the Hubble parameter at redshift $z$, and $\epsilon[\nu_{\rm obs} (1 + z')]$ is the proper volume emissivity of the emitted line.
With the assumption that the profile of each CO line is a delta function at the frequency $\nu_J$, we can express the emissivity as an integral of the host halo mass $M$:
\begin{equation}
 \epsilon(\nu, z) = \delta_D(\nu - \nu_J) (1 + z)^3 f_{\rm duty} \int_{M_{\rm min, CO}}^{\infty} dM \frac{dn}{dM} L_{\rm CO}(M,z)
\end{equation} 
Here, $L_{\rm CO} (M,z)$ is the specific luminosity of the CO line and it is assumed that a fraction $f_{\rm duty}$ of all haloes above a mass $M_{\rm min, CO}$ contribute to the CO emission.

With this, the specific intensity can be rewritten as:
\begin{equation}
I(\nu_{\rm obs}) = \frac{c}{4 \pi} \frac{1}{\nu_{\rm J} H(z_{\rm J})}  f_{\rm duty} \int_{M_{\rm min, CO}}^{\infty} dM \frac{dn}{dM} L_{\rm CO}(M,z)
\label{COspint}
\end{equation} 
The brightness temperature, $T_{\rm CO}$  can be derived from the specific intensity through the relation $I(\nu_{\rm obs}) = 2 k_B \nu_{\rm obs}^2 T_{\rm CO} /c^2$. 
Thus the expression for the brightness temperature becomes:
\begin{equation}
\langle T_{\rm CO} \rangle = \frac{c^3}{8 \pi}\frac{(1 + z_J)^2}{k_B \nu_J^3 H(z_J)} f_{\rm duty} \int_{M_{\rm min, CO}}^{\infty} dM \frac{dn}{dM} L_{\rm CO}(M,z)
\label{tco}
\end{equation} 
In order to derive the power spectrum to be observed in typical CO intensity mapping experiments, one also needs to model the clustering of the CO sources. In analogy with the methods for other species, e.g., neutral hydrogen intensity mapping, this can be done by weighting the dark matter halo bias by the CO luminosity-halo mass relation.
We thus have the expression for the clustering of CO sources:
\begin{equation}
 b_{\rm CO}(z) = \frac{\int_{M_{\rm min, CO}}^{\infty} dM (dn/dM) L_{\rm CO} (M,z) b(M,z)}{\int_{M_{\rm min, CO}}^{\infty} dM (dn/dM) L_{\rm CO} (M,z)}
\end{equation} 
{ where the $b(z)$ is the dark matter halo bias, e.g., given by \citet{scoccimarro2001}}.
The shot noise contribution to the power, due to the number of haloes, can now be expressed as:
\begin{equation}
 P_{\rm shot}(z) = \frac{1}{f_{\rm duty}}\frac{\int_{M_{\rm min, CO}}^{\infty} dM (dn/dM) L_{\rm CO} (M,z)^2}{\left(\int_{M_{\rm min, CO}}^{\infty} dM (dn/dM) L_{\rm CO} (M,z)\right)^2}
\end{equation} 
Given the above two expressions, we can express the signal (the power spectrum of the CO intensity fluctuations) as:
\begin{equation}
 P_{\rm CO}(k,z) = \langle T_{\rm CO} \rangle (z)^2 [b_{\rm CO}(z)^2 P_{\rm lin}(k,z) + P_{\rm shot}(z)]
\end{equation} 
as a function of $k$ at every redshift, from which we also have the power spectrum in logarithmic $k$-bins:
\begin{equation}
 \Delta_{k}^2(z) = \frac{k^3  P_{\rm CO}(k,z)^2}{2 \pi^2}
 \label{COpowspeclog}
\end{equation} 

In some studies, e.g. \citet{lidz2011}, $L_{\rm CO}(M)$ is simply modelled  as a linear relation: $L_{\rm CO}(M) = A_{\rm CO} M$, and $A_{\rm CO}$ is a proportionality constant. This reduces the expressions in Eqs. \ref{COspint} - \ref{COpowspeclog} to integrals over the dark matter halo mass alone.

\subsection{Models in the literature}
{{We thus see that one of the main astrophysical uncertainties in the measurement of the CO intensity power spectrum comes from the CO luminosity to host halo mass relation. 
Several approaches in the literature have been used to model this relation,  some of which are briefly summarized in \citet{li2015}. The astrophysical modelling typically requires (a) an SFR$-M$ relation and (b) an $L_{\rm CO} -$ SFR relation. The  various approaches towards modeling these are briefly described below (unless otherwise specified, the halo mass $M$ is in units of $M_{\odot}$, $L_{\rm CO}$ is in units of $L_{\odot}$ and the SFR is in units of $M_{\odot}$/yr):}}
\begin{enumerate}
\item In \citet{visbal2010}, the star formation rate is calculated as a function of halo mass as 
\begin{equation}
\mathrm{SFR} = 6.2 \times 10^{-11} \left(\frac{1+z}{3.5}\right)^{3/2} M
\end{equation}
and the CO luminosity is calculated from the SFR as:
 \begin{equation}
L_{\rm CO} = 3.7 \times 10^3 \ \mathrm{SFR}
\end{equation}
based on the observations of M82 in \citet{weiss2005}.
\item In \citet{pullen2013}, two models are described, Model A and Model B. In Model A, the star formation rate is calculated as:
\begin{equation}
\mathrm{SFR} = 1.2 \times 10^{-11}  M^{5/3}
\end{equation}
and the CO luminosity as
\begin{equation}
{L_{\rm CO}} =  3.2 \times 10^4 \ \mathrm{SFR}^{3/5}
\end{equation}
which is derived from the relations for $L_{\rm CO} - L_{\rm IR}$ \citep{daddi2010} and the $L_{\rm IR} - \rm{SFR}$ \citet{kennicutt1998}.
In Model B, empirical fits to the SFR are used, and the power spectra are multiplied by a rescaling factor  (which leads to about a factor 5 higher predicted brightness temperature at $z \sim 3$.)
\item In \citet{lidz2011}, the SFR is assumed proportional to the halo mass: 
\begin{equation}
\mathrm{SFR} = 1.7 \times 10^{-10} M
\end{equation} 
and it is also assumed proportional to the CO luminosity:
\begin{equation}
L_\mathrm{CO} = 3.2 \times 10^4 \ \rm{SFR}
\end{equation}
and the 5/3 power (assumed in the previous models) is replaced by unity for simplicity. 
\item In \citet{carilli2011}, the SFR assumed is that required to reionize the universe and keep it ionized, this is converted into an FIR luminosity by the relation \citep{kennicutt1998}:
\begin{equation}
L_{\rm FIR} = 1.1 \times 10^{10} \ \rm{SFR} 
\end{equation}
{{The FIR luminosity is, in turn}}, related to  the specific luminosity of the CO line, measured in units of K km/s pc$^2$ \citep[the median relation derived by][]{daddi2010}: 
\begin{equation}
 L'_{\rm CO} = 0.02 \ L_{\rm FIR} 
\end{equation} 
which can then be connected to the CO luminosity using 
\begin{equation}
L_{\rm CO} = 3.11 \times  10^{-11} \nu_r^3 L'_{\rm CO} 
\label{lsunlprime}                                           
     \end{equation} 
    { {where $\nu_r$ is the rest frequency of the transition under consideration.}}.

\item In \citet{righi2008}, the SFR- halo mass relation is derived following a merger history calculation with the extended Press-Schechter formalism of dark matter haloes \citep{lacey1993}. The SFR is then converted to CO luminosity using the scaling:
\begin{equation}
 L_{\rm CO} = 3.7 \times 10^3 \ \rm{SFR}
\end{equation}
from \citet{weiss2005}.

\item In \citet{gong2011}, the $L_{\rm CO}$ is modelled as a function of the halo mass at the reionization epoch ($z \sim 6-8$). It is fit using a function from the results of the semi-analytic modelling of \citet{obreschkow2009}):
\begin{equation}
L_{\rm CO} = L_0 \left(1 + \frac{M}{M_c}\right)^{-d} \left(\frac{M}{M_c}\right)^b
\end{equation}
with the values $L_0 = 4.3 \times 10^6, 6.2 \times 10^6, 4 \times 10^6 L_{\odot}$, $b = 2.4, 2.6, 2.8$ and $M_c = 3.5 \times 10^{11}, 3.0 \times 10^{11}, 2.0 \times 10^{11} M_{\odot}$  at redshifts 6, 7 and 8 respectively.

\item \citet{breysse2015} assume an SFR - CO relation derived from the results of \citet{carilli2013, pullen2013, lidz2011} based on the FIR - CO luminosity connection, which can be expressed as:
\begin{equation}
\mathrm{SFR} = 9.8 \times 10^{-18} \left(\frac{A_{\rm CO}}{2 \times 10^{-6}}\right) M^{5 b_{\rm CO}/3}
\end{equation}
where $L_{\rm CO} (M) = A_{\rm CO} M^{b_{\rm CO}}$, and the fiducial values are $A_{\rm CO} = 2 \times 10^{-6}, b_{\rm CO} = 1$.

\item \citet{mashian2015} use a large velocity gradient (LVG) modelling and an empirically determined star formation rate evolution to predict the power spectra corresponding to several CO transitions in the reionization era ($z \sim 6-10$). The star formation rate is modelled as a function of halo mass in the double power-law  form:
\begin{eqnarray}
\rm{SFR} &=& a_1 M^{b_1}, M\leq M_c; \nonumber \\
\rm{SFR} & = & a_2 M^{b_2}, M \geq M_c		
\end{eqnarray}
{{where $\{a_1, a_2, b_1, b_2\} = \{2.4 \times 10^{-17}, 1.1 \times 10^{-5}, 1.6, 0.6\}$ are the fitted parameters and the turnover occurs at the characteristic mass scale $M_c \approx 10^{11.6} M_{\odot}$.}}

\item \citet{li2015} use simulations of the galaxy-halo connection at redshifts 2.4-2.8 to model the intensity map and power spectrum of the CO (1-0) line at these redshifts.

\item \citet{fu2012} use different star formation prescriptions applied to semi-analytic models of galaxy formation to study the evolution of metals, atomic and molecular gas in galaxies including CO.

\end{enumerate}

These different approaches outlined above are found to lead roughly to an order of magnitude variation in the predicted CO luminosity - halo mass relation \citep[e.g.,][]{breysse2014, li2015}.

\section{Modelling the CO observables}
\label{sec:model}

{{
In this section, we begin by compiling the data available so far\footnote{We assume that the data and the errors quoted are representative. The method outlined, however, is sufficiently general as to be adapted to modifications and extensions to this data.}   in the context of the CO luminosity function, {{which is used in the subsequent analyses.}}

\begin{enumerate}

\item \citet{keres2003} use a sample of $\sim$ 300 galaxies from the FCRAO Extragalactic CO Survey \citep{young1995} at $z  = 0$ to derive a CO Luminosity Function (LF); and show that it is well fit by a Schechter function.

\item \citet{keating2016} provide constraints on the CO luminosity function at $z \sim 2.8$ by the measurement of the CO power spectrum in the COPSS (CO Power Spectrum Survey), this finds 
\begin{equation}
P_{\rm CO} = 3.0 \pm 1.3 \times 10^3 \mu {\rm{K}}^2 (h^{-1} {\rm{Mpc}}^3)
\end{equation}
at $z \sim 2.8$. This is combined with the data from direct detection efforts  to place constraints on the CO LF at $z \sim 3$, again assuming a Schechter form.

\item \citet{aravena2012} detect CO in (1-0) emission from a sample of four results from the Jansky Very Large Array (JVLA) survey at $z \sim 1.55$.

\item \citet{walter2014} use the results of a blind search in the Hubble Deep Field North (HDF-N) to place constraints on the CO luminosity function for the (1-0), (2-1) and (3-2) transitions at median redshifts 0.33, 1.52 and 2.75.
\end{enumerate}

The galaxy emission data \citep{keres2003}  suggest that the CO luminosity function, $\phi(L_{\rm CO})$ at $z\sim 0-3$ closely follows a Schechter form. Although the intensity mapping measurement \citep{keating2016} does not contain enough information to imply that the CO luminosity function at high redshifts is well fit by the Schechter function, the analysis suggests that it may a reasonable assumption in the light of the data available so far. Previous research in HI \citep{hpar2017, hpgk2016, hparaa2016},  suggests that this form of the luminosity function, reminiscent of a similar form for the HI (or stellar) mass function, leads to a distinct $L_{\rm CO}$-halo mass (or, equivalently, HI-halo mass) relation, when either derived directly \citep[e.g.,][]{hpar2017} or by abundance matching \citep[e.g.,][]{behroozi2013, moster2013, hpgk2016}. This assumes a monotonic relationship between the CO galaxies and the host haloes.
}}

{The data \citep[e.g., from COLD GASS,][]{saintonge2011, saintonge2011a, catinella2013} support a power law relation between the CO luminosity $L_{\rm CO}$ and the star formation rate, with an index $0.557 \sim 0.6$, which is consistent with theoretical predictions.} The existence of the star-forming main sequence \citep[SFMS, e.g.,][]{brinchmann2004, salim2007} which connects the star formation rate and stelllar mass of star-forming galaxies, supports a power-law relation between the SFR and the stellar mass ($M_{*}$) across a range of wavelengths and redshifts \citep[e.g.,][]{daddi2007}, such that we have $\mathrm{SFR} \propto M_{*}^{\beta}$ for both the $z > 1$ and $z < 1$ regimes. The relation is fairly tight and its normalization changes with redshift.

These findings can thus be combined to a power law form for the CO luminosity as a function of stellar mass. 
Further, using abundance matching of galaxies to dark matter haloes in simulations, the stellar mass - halo mass relation has been effectively modeled with a double power law behaviour \citep{moster2010, moster2013, behroozi2013}, and the evolution in the free parameters is fixed by matching to higher redshifts.

The above discussion offers support for a double power law behaviour for the $L_{\rm CO}- M$ relation (at redshift $z$) of the form:
\begin{equation}
L_{\rm CO} (M, z) = 2N(z) M [(M/M_1(z))^{-b(z)} + (M/M_1(z))^{y(z)}]^{-1}
\label{mosterco}
\end{equation}
with free parameters $M_1(z)$, $N(z)$, $b(z)$ and $y(z)$. These parameters are themselves composed of two terms, a constant term for $z \sim 0$, and an evolutionary component:
\begin{eqnarray}
\log M_1(z) &=& \log M_{10} + M_{11}z/(z + 1) \nonumber \\
N(z) &=& N_{10} + N_{11}z/(z + 1) \nonumber \\
b(z) &=& b_{10} + b_{11}z/(z + 1) \nonumber \\
y(z) &=& y_{10} +  y_{11}z/(z + 1)
\label{comoster}
\end{eqnarray}

\begin{figure}
 \begin{center}
 \includegraphics[scale=0.4, width = \columnwidth]{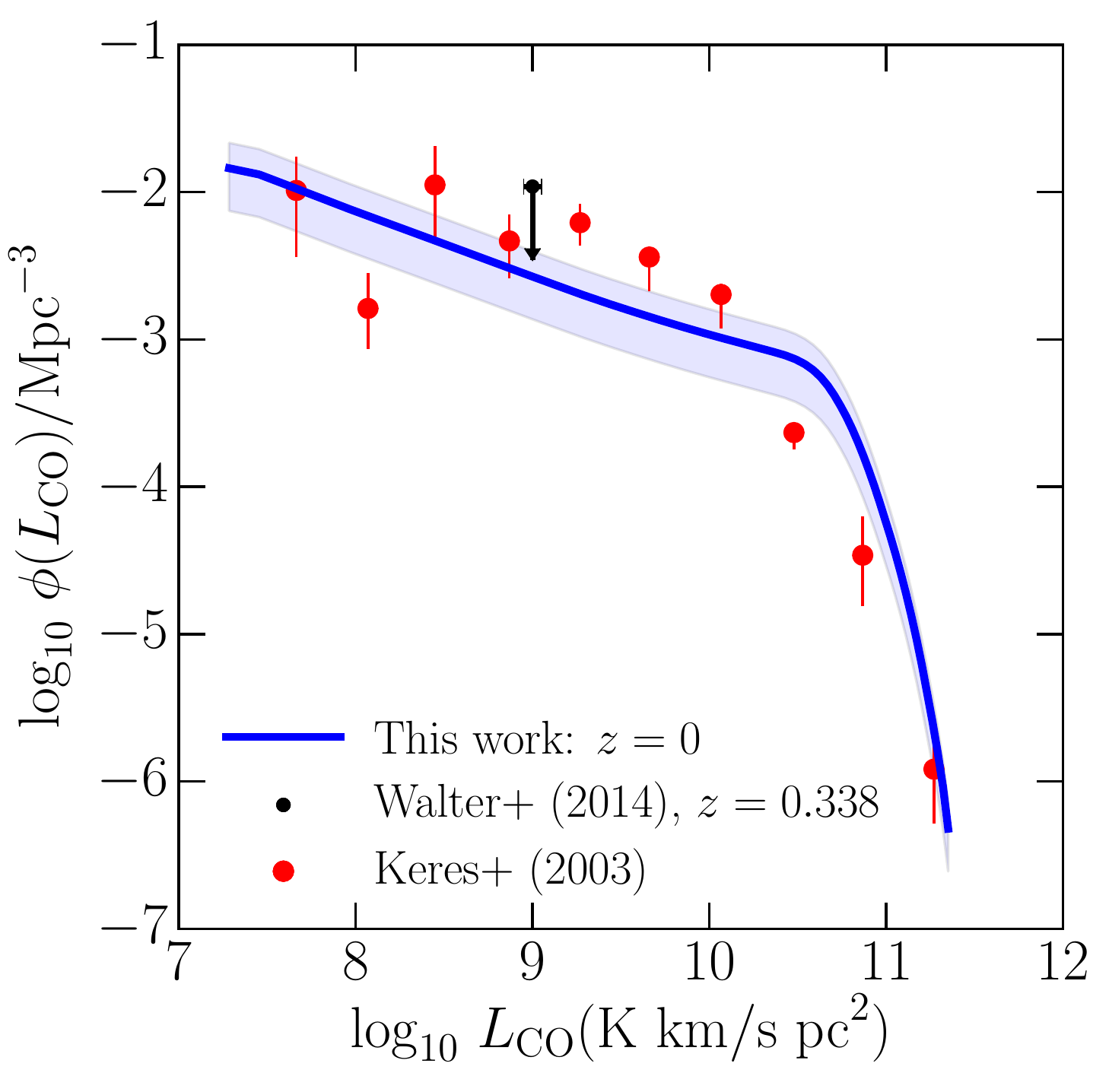}
 \end{center}
 \caption{The luminosity function of CO galaxies at $z \sim 0$. The red data points show the results from the FCRAO survey \citep{young1995, keres2003} at $z \sim 0$. The blue curve shows the derived luminosity function from the abundance matched best fitting parameters. The errors are indicated by the shaded regions. The black downward arrow shows the upper limit derived by \citet{walter2014} at $z \sim 0.34$.}
\label{fig:lumfunc}
\end{figure}

The CO luminosity function from
\citet{keres2003}  with a  sample of $\sim$ 300 galaxies (from the FCRAO Extragalactic CO Survey at $z  = 0$) is well fit by a Schechter function of the form: 
\begin{equation}
\phi (L_{\rm CO}) = \frac{dn}{d \log L_{\rm CO}} = (\ln 10) \ \rho ^* \left(\frac{L_{\rm CO}}{L*} \right)^{\alpha + 1} \exp-\left(\frac{L_{\rm CO}}{L*}\right)
\end{equation}
with the best-fit parameters: $\rho* = 0.00072 \pm 0.00035 \ \mathrm{Mpc}^{-3} \mathrm{mag}^{-1}, \alpha = -1.30 \pm 0.16$ and $L* = (1.0 \pm 0.2) \times 10^7$ Jy km/s Mpc$^{-2}$. The data points are shown in Fig. \ref{fig:lumfunc} in red along with the associated error bars. 

We use the Sheth-Tormen \citep[][]{sheth2002} prescription of for the dark matter halo mass function $dn/dM$. To recover the $L_{\rm CO} - M$ relation, we use the matching of the abundances of the halo mass function and the fitted CO
luminosity function, which can be expressed as \citep[e.g.,][]{vale2004}:
\begin{equation}
  \int_{M (L_{\rm CO})}^{\infty} \frac{dn}{ d \log_{10} M'} \ d \log_{10} M' = \int_{L_{\rm CO}}^{\infty} \phi(L_{\rm CO}) \ d \log_{10} L_{\rm CO}
  \label{eqn:abmatchco}
\end{equation}
In the above equation,  $dn / d \log_{10} M$ is the number density of dark matter haloes with logarithmic
masses between $\log_{10} M$ and $\log_{10} (M$ + $dM)$, and $\phi(L_{\rm CO})$ is the
corresponding number density of CO-luminous galaxies in logarithmic luminosity bins.  Solving
Equation~(\ref{eqn:abmatchco}) gives a relation between the CO luminosity
$L_{\rm CO}$ and the halo mass $M$.  This approach assumes
that there is a monotonic relationship between $L_{\rm CO}$ and $M$, which is reasonable in the light of the current data.

{{Abundance matching of the CO luminosity function obtained from \citet{keres2003}, to the dark matter halo mass function gives:
$M_{10} = (4.17 \pm 2.03) \times 10^{12} M_{\odot}, N_{10} = 0.0033 \pm 0.0016 \ \mathrm{K \ km/s  \ pc}^2 M_{\odot}^{-1}, b_{10} = 0.95 \pm 0.46, y_{10} = 0.66 \pm 0.32$. The data are binned into equally spaced logarithmic  bins in luminosity, between $\log L_{\rm CO} = 6$ and $\log L_{\rm CO} = 11$ in units of K km/s pc$^2$,  with a bin width = 0.1 dex. {\footnote{ {It can be checked (by increasing the bin width to 0.5 dex) that these results are not sensitive to reasonable changes in the number of bins within the quoted uncertainties.}}}  Errors on the parameters are estimated by a combination of the errors on the data and the fitting uncertainties. Plots of the (i) luminosity function data from the results of \citet{keres2003},  and (ii) the derived luminosity function from the abundance matched $L_{\rm CO} - M$ relation are shown in Fig. \ref{fig:lumfunc}. Also shown is the upper limit on the luminosity function measured by \citet{walter2014} at $z \sim 0.34$, for $L_{\rm CO} \sim 10^9$ K km/s pc$^2$.}}

{{The \citet{keating2016} measurement may chiefly sample the shot noise portion of the CO power spectrum. The second moment of the CO luminosity function thus measured, is combined with the data from direct detection studies \citep{decarli2014} and the absence of individual emitters within the COPSS dataset of $\geq 5 \sigma$  significance. This allows constraints on the CO luminosity function parameters at $z = 2.8$, assuming a Schechter functional form. These constraints are given by $\rho* = 1.3^{+0.6}_{-0.7} \times 10^{-3} \ L_{\odot}^{-1} \rm{Mpc^{-3}} \ \rm{mag}^{-1}$ and  $L_* = 4.5^{+1.4}_{-1.9} \times 10^{10} \  \rm{K \ km/s \ pc}^{−2}$,  with the prior $\alpha = -1.5 \pm 0.75$ which is based on the SFR function parameters from \citet{smit2012}.} 

{{The mean values and errors on the above Schechter function parameters  can now be used to fit the $L_{\rm CO} - M$ from \eq{comoster} at $z \sim 2.8$. This is done by matching the abundances of the halo mass function and the fitted CO luminosity function (with the associated errors) using \eq{eqn:abmatchco}. \footnote{For overall consistency,  we assume an identical bin range and number of bins as for the case of the $z = 0$ analysis.}}} With this, we obtain the redshift evolution parameters to be:
$M_{11} = -1.17 \pm 0.85, N_{11} = 0.04 \pm 0.03, b_{11} = 0.48 \pm 0.35, y_{11} = -0.33 \pm 0.24$. As in the previous case, the resultant errors are a combination of the fitting uncertainties and those from the data.  

The  resultant $L_{\rm CO} - M$ relations at redshifts 0, 1 , 2 and 3, along with their associated errors are shown in Fig. \ref{fig:hrlco}. Plotted for comparison at redshift $z \sim 2$ are the model predictions (which assume a linear $L_{\rm CO} - M$ relationship) from \citet{pullen2013} Model A and \citet{righi2008}. {{At redshift 3, the Model A prediction from \citet{pullen2013} is shown, as well as the constraint derived by COPSS II on $A_{\rm CO}(M)$, the coefficient of proportionality (assumed constant) between the CO luminosity and the host halo mass.}}

\begin{figure}
 \begin{center}
 \includegraphics[scale=0.3, width = \columnwidth]{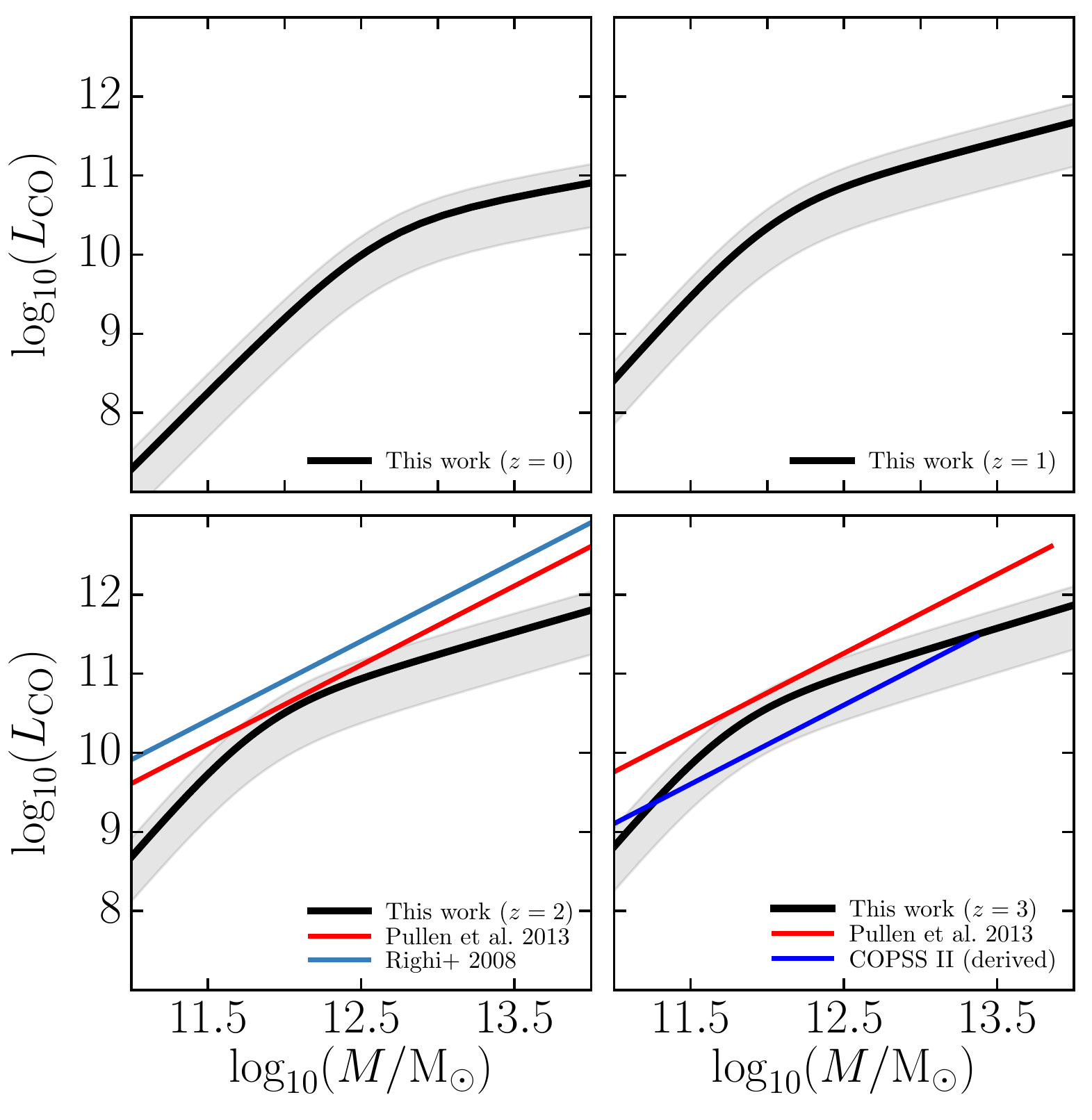} 
 \end{center}
 \caption{Best-fitting $L_{\rm CO} - M$ relation (with $L_{\rm CO}$ in units of K km/s pc$^2$) at $z = 0$, $z = 1$, $z = 2$ and $z = 3$, from the combined results of the low redshift \citet{keres2003} results and the higher redshift constraints from \citet{keating2016}. The associated errors are shown by the grey bands. Also shown in the panels are the estimates of \citet{pullen2013} Model A and \citet{righi2008} at $z \sim 2-3$. At $z \sim3$, the {derived estimate from the COPSS II data \citep{keating2016} which assumes a constant $A_{\rm CO}(M)$ is also shown.}}
\label{fig:hrlco}
\end{figure}

\section{Comparison to data}
\label{sec:compare}

We have seen (Fig. \ref{fig:lumfunc})   that the predicted luminosity function at $z \sim 0$ is consistent by construction with the \citet{keres2003} data, and is also consistent with the upper limit from \citet{walter2014}. 

\citet{aravena2012} use the results from a Jansky Very Large Array (JVLA) survey for CO 1-0 line emission from a candidate cluster at $z \sim 1.55$, targeting four galaxies in the redshift range 1.47 to 1.59. Previous simulations were found to somewhat underestimate the number of CO galaxies detected at this redshift.
In Fig. \ref{fig:red1_5} is plotted the model luminosity function with its associated error at $z \sim 1.5$, compared to the findings of \citet{aravena2012}. {{The data point shows the result for all the four galaxies in the sample, and is consistent with the model predictions.}} This is a consequence of the fact that the present model is also anchored to the high-redshift data [the $z \sim 2.75$ measurements from \citet{keating2016}]. Also plotted in Fig. \ref{fig:red1_5}  are the observational results from \citet{walter2014}, in the redshift range $1.01 < z < 1.89$ (median redshift 1.52)  from a blind search in the Hubble Deep Field North (HDF-N). 

In Fig. \ref{fig:red2_75} are plotted the results from this work at $z \sim 2.75$ with the associated error bars,  and for comparison, the results from COPSS \citep{keating2016} which is consistent by construction. Also shown are the results from \citet{walter2014} with the median redshift $z = 2.75$.
Our findings are consistent with the fact that the \citet{walter2014} and the COPSS \citep{keating2016} results are in agreement, as also noted by \citet{keating2016}. The present model predictions at these redshifts are also somewhat higher than those estimated by previous simulations. 

The molecular hydrogen abundance can be constrained using estimates for the CO-to H$_2$ conversion factor and the total luminosity of the CO galaxies. With a typical value of $\alpha = 4.3$, the cosmic hydrogen abundance is found to be $\rho_{\rm{H}_2} \approx 10^8 M_{\odot} \ \mathrm{Mpc}^{-3}$ at $z \sim 3$, in good agreement with the results from data and theoretical models \citep{obreschkow2009, lagos2011, sargent2014, popping2015, walter2014}.

\begin{figure}
 \begin{center}
 \includegraphics[scale=0.4, width = \columnwidth]{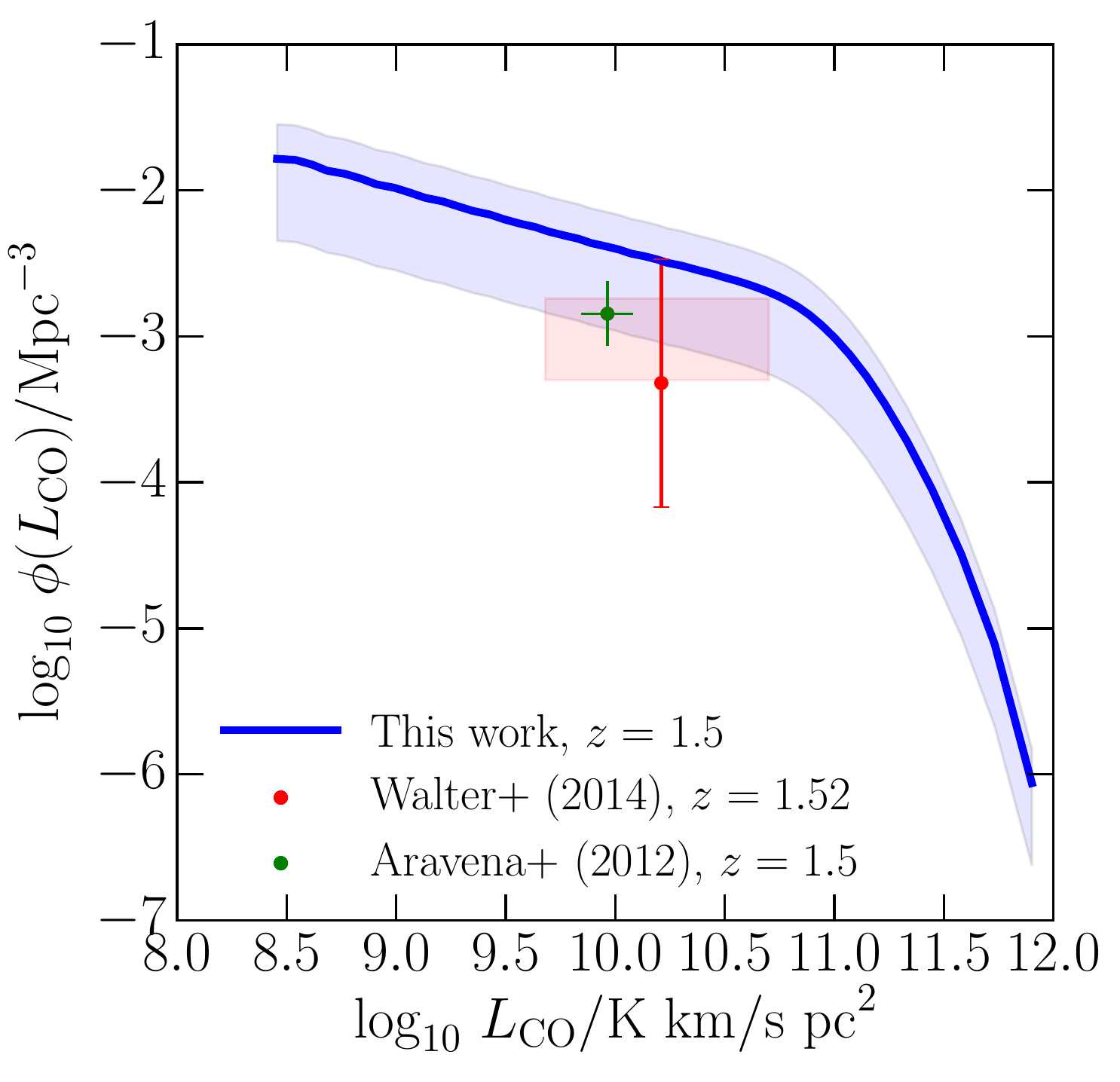}
 \end{center}
 \caption{The CO luminosity function at $z \sim 1.5$ with the associated error shown by the shaded bands. Also plotted for comparison are the results from \citet{aravena2012} and \citet{walter2014} at this redshift.}
\label{fig:red1_5}
\end{figure}

\begin{figure}
 \begin{center}
 \includegraphics[scale=0.4, width = \columnwidth]{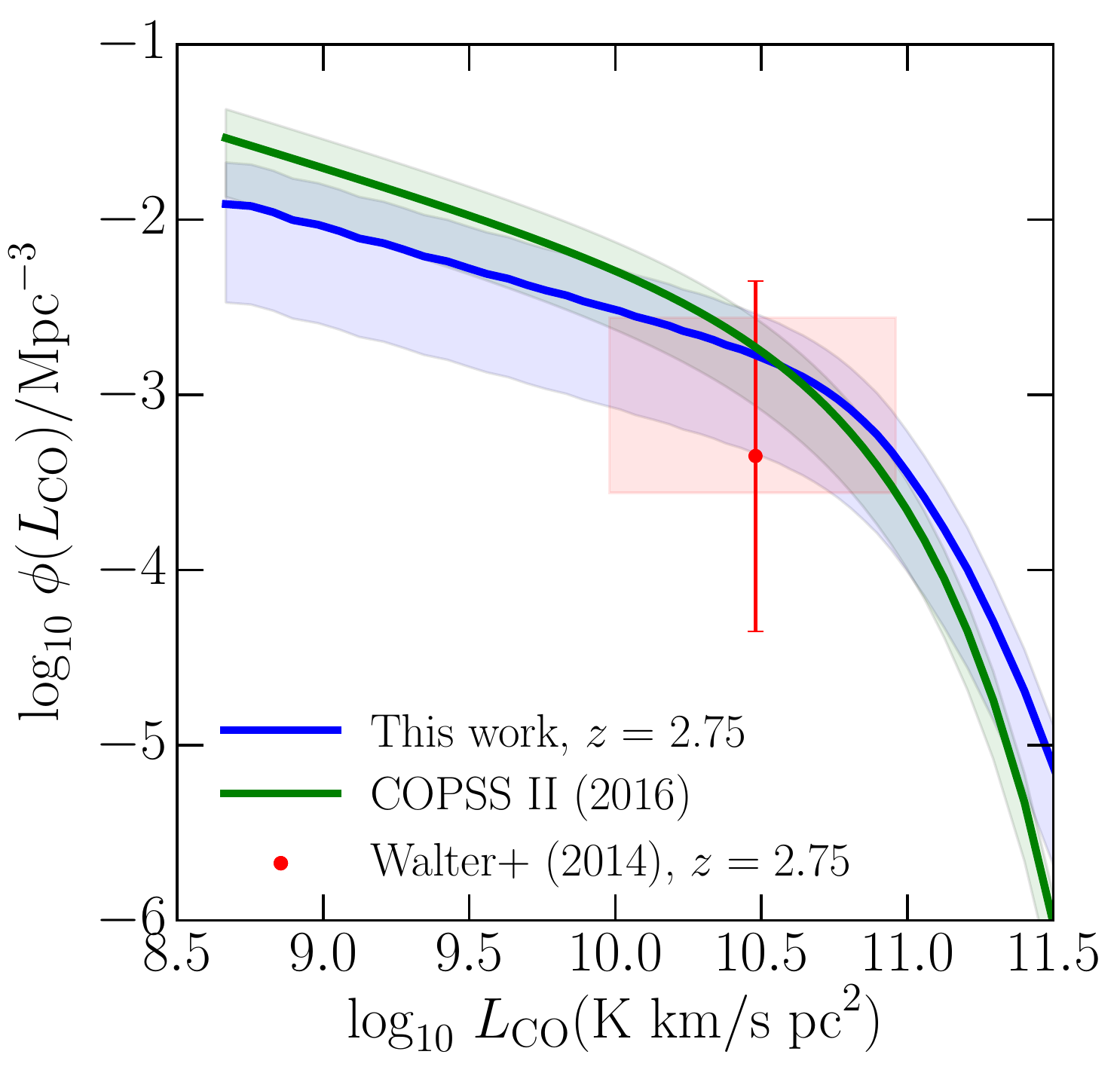}
 \end{center}
 \caption{The CO luminosity function at $z \sim 2.75$. The associated errors are shown by the shaded bands. Also plotted for comparison are the results from \citet[COPSS;][]{keating2016} and \citet{walter2014} at this redshift.}
\label{fig:red2_75}
\end{figure}

Using the predicted $L_{\rm CO} - M$ relation, we can estimate the magnitude and uncertainties of  the CO temperature evolution and the intensity mapping power spectrum. We focus here on the 1-0 transition; analogous methods can be applied to the higher transitions as well. 

The predicted $T_{\rm CO}$ at various redshifts from the present model, using \eq{tco} is shown in Fig. \ref{fig:tcored3} by the blue dashed line.\footnote{This is calculated following the conventions in \citet{breysse2014}, Eq. (2.5) to enable ease of comparison with the compiled results in that work.} We assume fiducial values of $f_{\rm duty} = 0.1$ and $M_{\rm min, CO} = 10^9 h^{-1} M_{\odot}$ in this plot. The shaded area indicates the model uncertainty.\footnote{Note that the  uncertainties in $f_{\rm duty}$ and $M_{\rm min, CO}$ are not included in the error band, which therefore represents a lower limit.} The figure also shows the predictions from various other models in the literature, compiled in \citet{breysse2014} at $z \sim 3$. It can be seen that the model predictions are consistent with the results of \citet{righi2008} and \citet{pullen2013} Model A in the previous literature, but below the Model B in \citet{pullen2013}. {{This is as expected since the present model is matched to the results of \citet{keating2016}, whose data are also found to be below Model B of \citet{pullen2013}.}} The model is marginally consistent with the results of \citet{visbal2010}.
\begin{figure}
 \begin{center}
 \includegraphics[scale=0.4, width = \columnwidth]{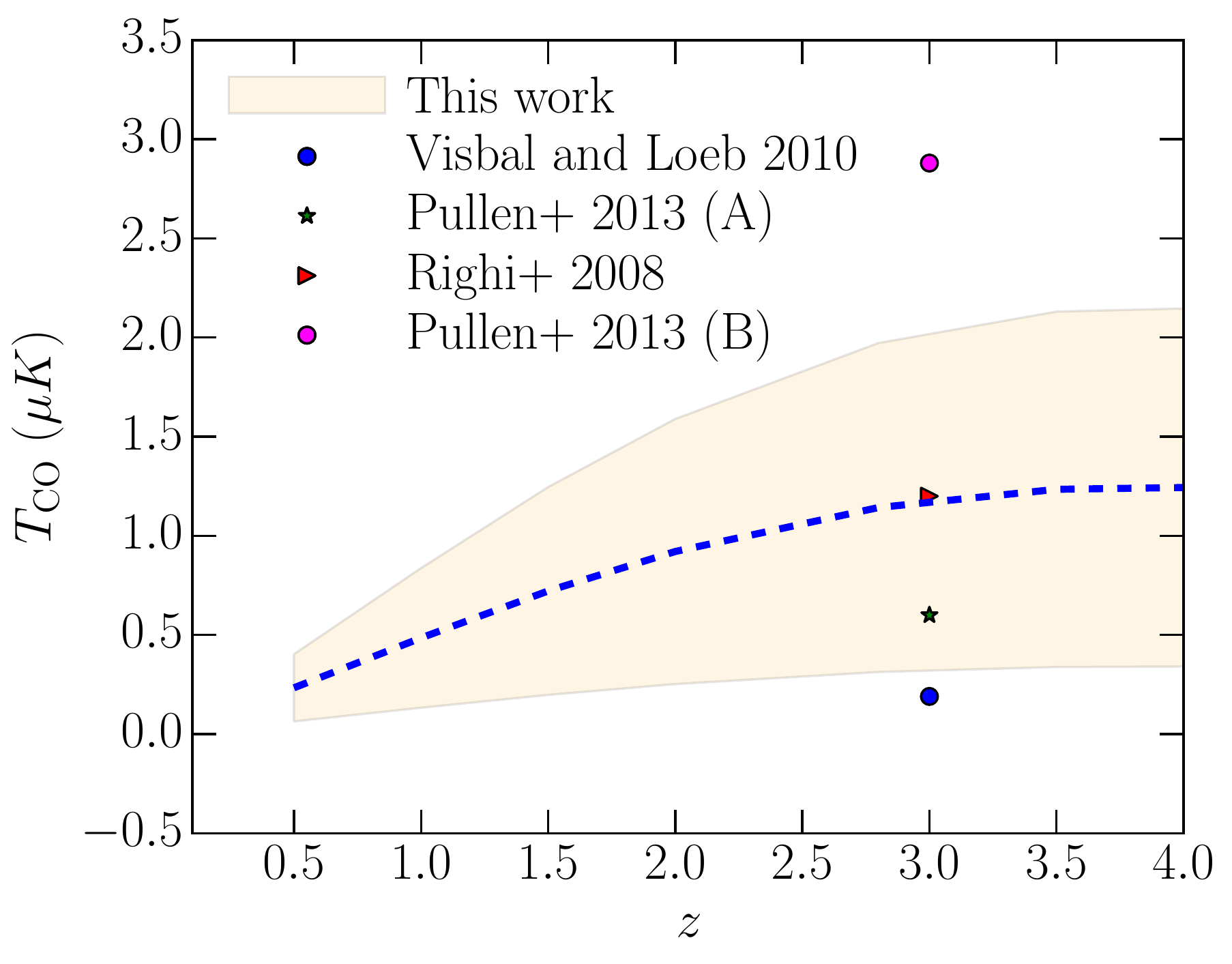}
 \end{center}
 \caption{The best-fitting evolution of the mean $T_{\rm CO}$ (dashed blue curve) with orange error band. Results from the theoretical predictions of \citet{visbal2010, pullen2013, righi2008} at $z \sim 3$ are also shown for comparison (compiled in \citet{breysse2014}).}
\label{fig:tcored3}
\end{figure}

Finally, we can use the model predictions to compute the CO intensity mapping power spectrum using \eq{COpowspeclog}. {{This calculation depends on the values of the minimum host halo mass, $M_{\rm min}$, and also the duty cycle factor ($f_{\rm duty}$). The minimum mass is assumed to be $M_{\rm min, CO} = 10^9 h^{-1} M_{\odot}$ throughout. The power spectra (in units of $\mu K^2$) computed with two fiducial values of $f_{\rm duty}$: 0.1 and 1, are shown in the top panel of Fig. \ref{fig:powspec2}. These are compared with the model of \citet{li2015} at the midpoint of the redshift range probed by the COMAP experiment ($z \sim 2.4 - 2.8$) . The COMAP experiment sensitivity is also indicated on this panel by the red curve.  

Although tight constraints on $f_{\rm duty}$ are difficult with the current data, most of the observational evidence suggests (and uses) a value of $f_{\rm duty}$  close to unity \citep{keating2016}. The bottom panel plots this along with the COPSS II data above the noise limit (black points).}}

\begin{figure}
 \begin{center}
 \includegraphics[scale=0.4, width = \columnwidth]{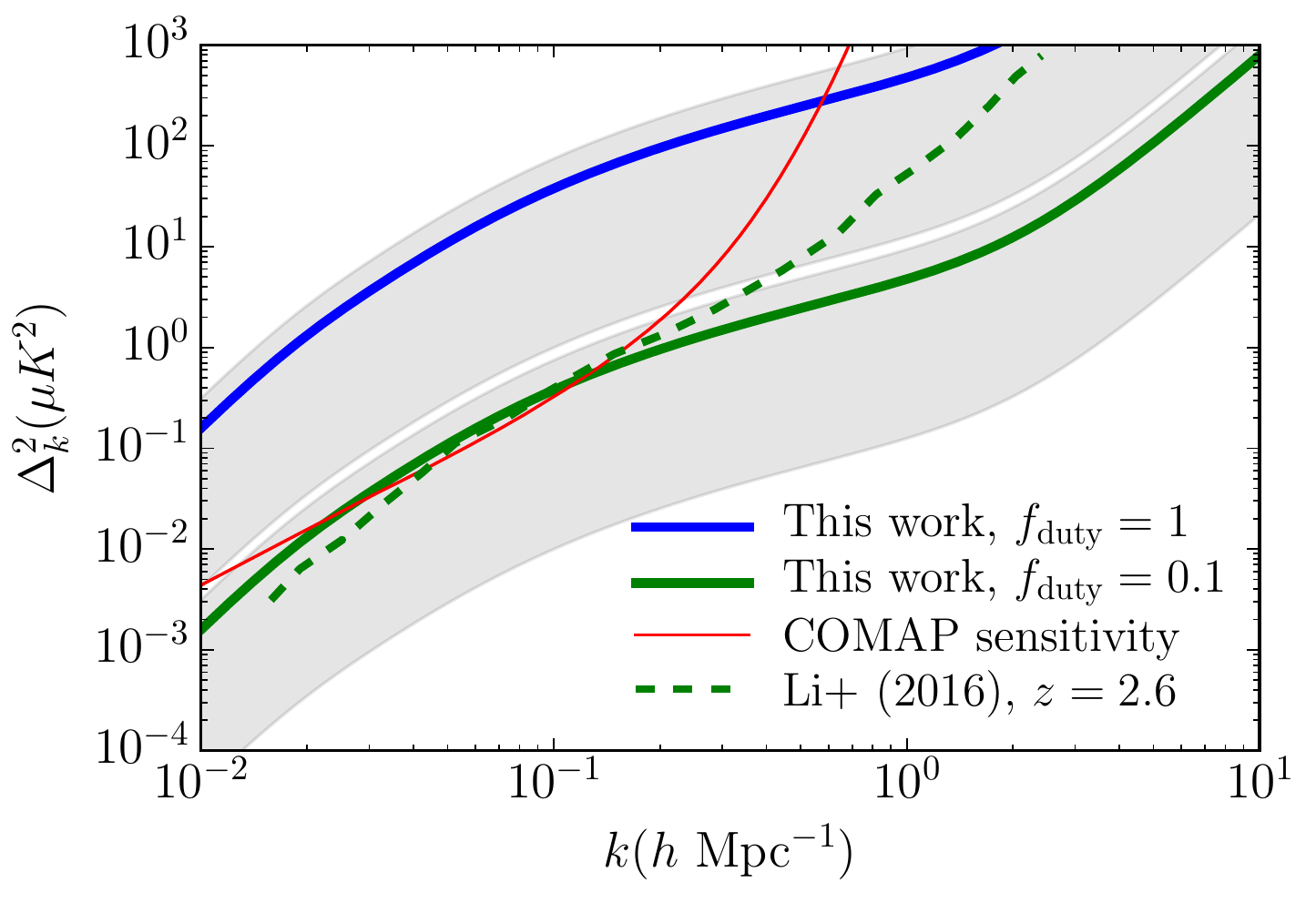} \includegraphics[scale=0.4, width = \columnwidth]{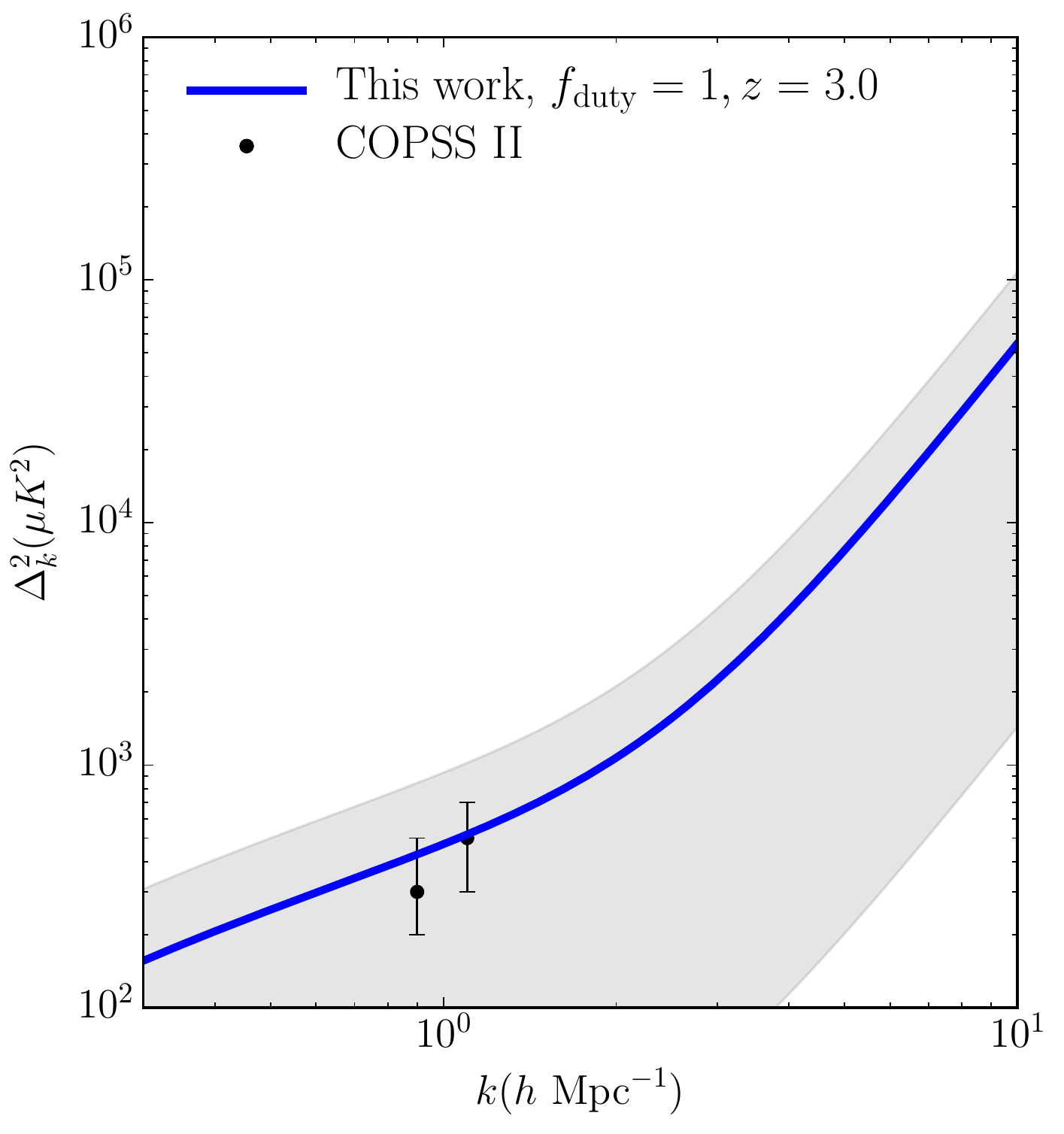}
 \end{center}
 \caption{{{ The predicted best-fitting CO power spectrum,  $\Delta^2_{k}$ at redshifts $z = 2.6$ (top panel) and $z = 3.0$ (bottom panel). The associated errors are shown by the shaded bands. In the top panel, two values of $f_{\rm duty}$ are considered: 0.1 and 1, and the simulation from \citet{li2015} at the COMAP redshifts 2.4-2.8  is plotted as the dashed green curve. The COMAP sensitivity is shown in this panel as the red line. The bottom panel shows the power spectrum with an assumed 100\% duty cycle, along with the results of COPSS II above the noise limit (shown by the black data points).}}}
\label{fig:powspec2}
\end{figure}

\section{Discussion and outlook}
\label{sec:conc}
In this paper, we have compiled the recent data in the field of carbon monoxide (CO) 1-0 emission line observations at low and intermediate redshifts. Here, we briefly summarize our results, discuss the scope of the technique and outline the possibilities for future work.

\begin{enumerate}

\item We have used the data at low redshifts to constrain the evolution of a parametric $L_{\rm CO}$ - halo mass  relation derived empirically. Given that the CO luminosity functions are well fit by the Schechter form, it is reasonable to expect the derived CO - halo mass relation to be modeled analogously to the stellar mass - halo mass relation \citep[SHM;][]{behroozi2013, moster2013}. 

\item This assumes a one-to-one-relationship between the host haloes and the CO-luminous galaxies, and also  the completeness of the sample(s) under consideration. A caveat to the technique is that the halo mass function assumed is theoretical, and the assumption of matching the most massive haloes is involved. However, being completely empirical, this approach is free from the modelling uncertainties present in simulations, and at the same time is complementary to those studies. Extensions to this framework may be possible with the help  of future data and comparison to high-resolution hydrodynamical simulations.

\item The evolution of the free parameters is determined from matching the available constraints at higher redshift ($z \sim 3$) from intensity mapping. The resulting CO - halo mass relation is found to be consistent with most predictions from previous literature. It is also consistent with the results of surveys at intermediate redshifts \citep{aravena2012,walter2014}. The associated errors not only encompass the uncertainties in the data, but also a range of uncertainties in the theoretical modelling. Thus, the fitting forms and errors contain the available theoretical and observational constraints on the intensity mapping power spectrum. 

\item Using the empirically determined $L_{\rm CO} - M$ relation with fiducial values of the minimum mass $M_{\rm min}$ and duty fraction $f_{\rm duty}$, one can predict the evolution of the integrated brightness temperature of the CO emission, $T_{\rm CO} (z)$ and the power spectrum $P_{\rm CO} (k,z)$ as a function of scale and redshift. These predictions are in, turn, consistent with the results of simulations in the literature. Table \ref{table:final} summarizes the fitting functions for the $L_{\rm CO} - M$ relation derived using the present approach.

\end{enumerate}

\begin{table}
\centering
\caption{Summary of the best-fitting $L_{\rm CO} - M$ relation across $z \sim 0-3$,  and the free parameters involved. The $L_{\rm CO}$ is in units of  K km/s pc$^2$ and all masses are in units of $M_{\odot}$.}
\label{table:final}
\begin{tabular}{|c|}
\hline
\\
$L_{\rm CO} (M, z) = 2N(z) M [(M/M_1(z))^{-b(z)} + (M/M_1(z))^{y(z)}]^{-1}$;\\ 
\\
$\log M_1(z) = \log M_{10} + M_{11}z/(z + 1)$
 \\
 \\
$N(z) = N_{10} + N_{11}z/(z + 1)$
 \\
 \\
$b(z) = b_{10} + b_{11}z/(z + 1)$
 \\
 \\
$y(z) = y_{10} +  y_{11}z/(z + 1)$
 \\
\\
\hline 
\\
$M_{10} = (4.17 \pm 2.03) \times 10^{12} \ M_{\odot}$                                                                                                                                                                                                                                                                                                                                                                       ;\ $M_{11} = -1.17 \pm 0.85 $                                                                                                                                                                                                                                                                                                                                                                               \\
\\ $N_{10} = 0.0033 \pm 0.0016$                                                                                                                                                                                                                                                                                                                                                                         ; \ $N_{11} = 0.04 \pm 0.03$                                                                                                                                                                                                                                                                                                                                                                              
 \\
\\$b_{10} = 0.95 \pm 0.46$                                                                                                                                                                                                                                                                                                                                                                       ;\ $b_{11} = 0.48 \pm 0.35$                                                                                                                                                                                                                                                                                                                                                                              
 \\
\\$y_{10} = 0.66 \pm 0.32$                                                                                                                                                                                                                                                                                                                                                                       ;\ $y_{11} = -0.33 \pm 0.24$                                                                                                                                                                                                                                                                                                                                                                              
 \\

\hline
\end{tabular}
\end{table}

Tighter constraints on the power spectrum might be possible with new measurements from future detections (at low and intermediate redshifts)  from a large sample of galaxies, {e.g., with the ALMA Spectroscopic Survey in the Hubble Ultra Deep Field (ASPECS) survey \citep{walter2016}} and intensity mapping with facilities like the COMAP and the Y. T. Lee Array \citep[YTLA;][]{ho2009}. Likewise, with the availability of new data, the model can be extended by, e.g., introducing merger histories and more accurate treatments of star formation \citep[as done for the stellar-halo mass in, e.g.][]{moster2010}, and also to account for the turnover in the star-formation rate density beyond $z \sim 3$. 

It would be interesting to investigate the possibility of empirically constraining the $f_{\rm duty}$ factor and connecting it to physically motivated duty cycles used in models of the UV luminosity function \citep[e.g.,][]{tacchella2013}. With high-redshift data, the approach may be connected to the existing frameworks for modelling CO at close to the reionization epoch \citep[$z \sim 6-10$; as done in, e.g.,][]{mashian2015, gong2011}. Recently, a large sample of local CO-emitting galaxies has been compiled by \citet{boselli2014}, which may be useful to constrain the CO density profiles and enable a more detailed characterization of the 1-halo term involved in the clustering \citep[as done for HI in, e.g.,][]{hparaa2016}. Similarly, it would be useful to extend this approach towards the abundances of other molecules like \textsc{c ii} \citep[which has been modelled for the reionization epoch in, e.g.,][]{gong2012} and thereby facilitate the study of intensity mapping cross-correlations. 

\section*{Acknowledgements}
{I thank Kieran Cleary, Anthony Readhead, Adam Amara, Benny Trakhtenbrot, Tony Li and members of the COMAP (CO Mapping Array Pathfinder) collaboration for insightful discussions. I thank Jasjeet Bagla, Patrick Breysse, Kieran Cleary, Robert Feldmann, Girish Kulkarni, Anthony Pullen, Nirupam Roy, Sandro Tacchella and Livia Vallini for useful comments on the manuscript, and the referee for a detailed and helpful report.} My research is supported by the Tomalla Foundation.

\bibliographystyle{mnras}
\bibliography{mybib}

\bsp
\label{lastpage}
\end{document}